\documentclass[]{aa}

\usepackage{amsmath}
\usepackage{graphicx}
\usepackage{natbib}
\usepackage{bm}
\bibpunct{(}{)}{;}{a}{}{,}

\begin{document}

\title{Image Reconstruction with Analytical Point Spread Functions}
\author{A. Asensio Ramos\inst{1,2} \& A. L\'opez Ariste\inst{3}}

\offprints{aasensio@iac.es}

\institute{
 Instituto de Astrof\'{\i}sica de Canarias, 38205, La Laguna, Tenerife, Spain; \email{aasensio@iac.es}
\and
Departamento de Astrof\'{\i}sica, Universidad de La Laguna, E-38205 La Laguna, Tenerife, Spain
\and
THEMIS, CNRS UPS 853, c/v\'{\i}a L\'actea s/n. 38200. La  Laguna,
Tenerife, Spain; \email{arturo@themis.iac.es}
}
\date{Received ; Accepted}
\titlerunning{Image Reconstruction with Analytical PSFs}
\authorrunning{Asensio Ramos \& L\'opez Ariste}
% 
% 
% % abstract cannot exceed 300 words
\abstract
{The image degradation produced by atmospheric turbulence and optical aberrations is 
usually alleviated using post-facto image reconstruction techniques, even when observing
with adaptive optics systems.}
{These techniques rely on the development of the wavefront using Zernike functions and the
non-linear optimization of a certain metric. The resulting optimization procedure is  
computationally heavy. Our aim is to alleviate this computationally burden.}
{To this aim, we generalize the recently developed extended Zernike-Nijboer theory to carry out the 
analytical integration of the Fresnel integral and
present a natural basis set for the development of the point spread function in case the
wavefront is described using Zernike functions.}
{We present a linear expansion of the point spread function in terms of analytic
functions which, additionally, takes defocusing into account in a natural way. This expansion
is used to develop a very fast phase-diversity reconstruction technique which is demonstrated
through some applications.}
{This suggest that the linear expansion of the point spread 
function can be applied to accelerate other reconstruction techniques in use presently and based on
blind deconvolution.}

\maketitle

%%%%%%%%%%%%%%%%%%%%%%%%%%%%%%%%%%%%%%%%
%%%%%%%%%%%%%%%%%%%%%%%%%%%%%%%%%%%%%%%%
% INTRODUCTION
%%%%%%%%%%%%%%%%%%%%%%%%%%%%%%%%%%%%%%%%
%%%%%%%%%%%%%%%%%%%%%%%%%%%%%%%%%%%%%%%%
\section{Introduction}
Atmospheric turbulence degrades astronomical images by introducing aberrations
in the wavefront. During the last years, 
functional adaptive optics systems have been developed with the aim of
partially cancelling these aberrations and generating a diffraction limited
image of any astronomical object \citep[e.g.,][]{beckers_ao93}. The development
of such systems has been particularly challenging for solar telescopes because
the reference object (the solar surface) has spatial structure.
In spite of the great success of adaptive optics systems \citep{rimmele_ao00,scharmer_ao00}, the limited number of 
degrees of freedom of the deformable mirror and the speed at which it has to
work, especially in solar observations, impedes a full correction of the wavefront. 
Therefore, it has become customary in solar imaging to apply post-facto reconstruction techniques
to the observed images to improve the wavefront correction. The advantage of
these techniques is that the time limitation is much less restrictive since
they do not need to work in real-time. Hence, one can use lots of computation power
to reconstruct the images. One of the methods considered first for the
reconstruction of solar images was based on speckle techniques \citep{vonderluhe93,vonderluhe94}
which produced images of extraordinary quality. However, it suffers from two fundamental problems.
On the one hand, the number of images that one needs to acquire is large (although
comparable with the number of images needed in complex blind deconvolution algorithms) 
because the missing information 
on the wave phases is statistically estimated from a succession of many short exposures. On the other
hand, it relies on a theoretical description of the effect of the atmospheric turbulence
and adaptive optics on the image \citep{woger07}, which might be inaccurate. However, recent efforts have shown that
almost real-time processing \citep{denker01} and photometric precision \citep{woger08}
can be achieved routinely.

A complementary approach to speckle reconstruction appeared
with the systematic application of techniques based on phase diversity \citep[e.g.,][]{lofdahl94,lofdahl98},
developed some years before by \cite{gonsalves79} and later extended by \cite{paxman92}. These
techniques rely on a maximum likelihood simultaneous estimation of the wavefront and the 
original image (before being affected by  atmospheric+telescope aberrations) using a reference 
image and a second one that contains exactly the same aberrations as the reference one plus
a known artificially-induced aberration. Because of the mechanical simplicity, a defocus typically
constitutes the added aberration. The main drawback of the phase diversity technique
is the heavy computational effort needed to compute the maximum likelihood solution. Many
Fourier transforms have to be used to estimate the point spread function (PSF) from
the wavefront at the pupil plane.

Even more computationally expensive is the blind deconvolution technique developed by 
\cite{vannoort05}. This technique, termed multi-object multi-frame blind deconvolution (MOMFBD),
is able to reconstruct estimates of the object and the wavefront from the observation of
different objects (same location in the Sun observed at different wavelengths) with
some constraints, which are essentially used to augment the amount of information introduced in 
the likelihood function. As a consequence, the estimation of the wavefront and the
original image is less dominated by noise and hence more stable. In close analogy with phase diversity, this enhancement
in stability comes at the price of a higher computational burden since many Fourier transforms
have to be computed for the maximization of the likelihood.
In conclusion, all these techniques suffer from time-consuming computations. As modern solar telescopes 
increase their flux of high quality data, observers are condemned to struggle with the bottleneck of reconstruction algorithms.

During the 1940's, \cite{nijboer42} made it possible to estimate the PSF of an aberrated 
optical device in terms of the generalized complex pupil function. The resulting integral expressions were
very involved and remained impossible to evaluate until recent years. What is presently
known as the extended Nijboer-Zernike theory \citep[ENZ;][]{janssen_enz02,braat_enz02} presents
analytical expressions for the complex field at the focal volume that depend on the aberration through the
coefficients of the Zernike expansion of the wavefront. The expressions are closed in the sense
that they only depend on computable functions and the coefficients of the Zernike expansion.

Our purpose in this paper is to extend the ENZ theory to arbitrary aberrations exploiting mathematical 
properties of the Zernike functions recently found. As a consequence, we find a general expression for the
PSF in the focal volume as a linear combination of given basis functions. Starting from this expression,
we propose to use it in a phase diversity reconstruction algorithm. Due to the analytical
character of the PSF, we are able to avoid the calculation of Fourier transforms during the
iterative process, thus inducing an important gain in computational time. 
This shall be our fundamental result: that through this formalism we can drastically diminish 
the computational burden of reconstruction algorithms and therefore the time needed to perform the 
reconstruction. We illustrate it using phase diversity. But in principle 
any reconstruction algorithm can benefit from our formalism, since we only change
the way in which the PSF is described.

The outline of the paper is the following. Section \ref{sec:wavefront_expansion} gives a description
of the wavefront expansion in Zernike functions and how the field distribution can be
analytically obtained for general aberrations. The expansion of the point spread function
is presented in \S\ref{sec:psf} with an statistical analysis of that expansion for atmospheric
Kolmogorov turbulence in \S\ref{sec:statistical}. Section \ref{sec:phase_diversity} presents
the application of the formalism to the fast restoration of images using phase-diversity.
Finally, the conclusions are presented in \S\ref{sec:conclusions}.

%%%%%%%%%%%%%%%%%%%%%%%%%%%%%%%%%%%%%%%%
%%%%%%%%%%%%%%%%%%%%%%%%%%%%%%%%%%%%%%%%
% Analytical Point Spread Function
%%%%%%%%%%%%%%%%%%%%%%%%%%%%%%%%%%%%%%%%
%%%%%%%%%%%%%%%%%%%%%%%%%%%%%%%%%%%%%%%%
\section{Wavefront expansion}
\label{sec:wavefront_expansion}
Let $P(\rho,\theta)$ represent the wavefront of a point source beam in a pupil
plane, with $\rho$ and $\theta$ the polar coordinates of the unit
disk in that plane. The field
distribution at an image plane can be written in the Fresnel framework as \citep{born_wolf80}:
\begin{equation}
 U(r,\varphi;f)=\frac{1}{\pi}\int_0^1\int_0^{2\pi} \rho d\rho d\theta
e^{if\rho^2} P(\rho,\theta) e^{2\pi i\rho r \cos(\theta-\varphi)},
\label{eq:fresnel}
\end{equation}
which is, in general, a complex quantity.
In that expression we have defined a coordinate system $(r,\varphi;f)$ over the
image space, where $(r,\varphi)$ are angular coordinates in the focal plane and
$f$ is the distance to that focal plane. We define the focal plane as the plane
where the optical system would form the image of the object in the absence of any aberration.
We implicitly assume an optical system with numerical aperture $NA\ll 1$, without loss of 
generality for solar instruments. The
$(r,\varphi;f)$ coordinates can be put in relationship with the geometrical
coordinates $(X,Y,Z)$ of the optical system by means of the following
relationships
\begin{eqnarray}
x&=&X\frac{2\pi(NA)}{\lambda}, \quad y=Y\frac{2\pi(NA)}{\lambda}, \quad f=Z\frac{\pi(NA)^2}{\lambda} \nonumber \\
r&= &\sqrt{x^2+y^2}, \quad \tan \varphi =\frac{y}{x}, \nonumber \\
\end{eqnarray}
where $\lambda$ is the wavelength of the light.

%%%%%%%%%%%%%%%%%%%%%%%%%%
%%%%%%%%%%%%%%%%%%%%%%%%%%
\subsection{Wavefront expansion}
In order to solve the integral in Eq. (\ref{eq:fresnel}), we have to provide a useful
description of the wavefront $P(\rho,\theta)$. In general the wavefront is
also a complex quantity which is usually represented in polar form: the modulation in amplitude
is represented by $A(\rho,\theta)$ while the phase modulation is determined by $\Phi(\rho,\theta)$. 
If the pupil plane is fully transmissive (or perfectly reflective) we can let $A(\rho,\theta)=1$ so that:
\begin{equation}
P(\rho,\theta)= A(\rho,\theta) e^{i \Phi(\rho,\theta)}=e^{i \Phi(\rho,\theta)}.
\end{equation} 
This is the case of a clear aperture telescope with the wavefront
modified by atmospheric turbulence. For the sake of simplicity, we assume this to be the case, since
it produces simpler mathematical expressions that can be
related more transparently to previous results. An annular pupil would only require a
change of the integration limits in Eq. (\ref{eq:fresnel}), thus modifying accordingly
the final result.

It is customary (and for many reasons interesting) to express the phase modulation
of the wavefront in terms of the Zernike functions \citep{noll76}. 
The Zernike functions are labeled by two quantum numbers: a principal number $n$
and an azimuthal number $m$ fulfilling that $|m| \leq n$ and that $n-|m|$ is even. Their definition
is:
\begin{eqnarray} 
Z_n^m(\rho,\theta) &=& R_n^m(\rho) \cos (m\theta) \\
Z_n^{-m}(\rho,\theta) &=& R_n^m(\rho) \sin (m\theta) \\
Z_n^0(\rho,\theta) &=& R_n^0(\rho),
\end{eqnarray}
where $R_n^m(\rho)$ is a Zernike polynomial \citep[e.g.,][]{born_wolf80}. 
In order to simplify mathematical manipulations, \cite{noll76} introduced a
useful single ordering index $j$, with perfect
correspondence with pairs $(n,m)$ so that higher values of $j$ represent higher
aberrations with smaller probability amplitudes in a Kolmogorov turbulent atmosphere. We will
use the two notations indistinctly. The reader should be aware that any 
pair ($n$,$m$) appearing explicitly in an expression has to be in agreement with 
the corresponding Noll index $j$.
Since Zernike
functions constitute a family of orthogonal functions in the unit circle, any phase aberration can be written 
as a linear expansion:
\begin{equation}
\label{eq:phase_aberration}
\Phi(\rho,\theta) = \sum_j a_j Z_j(\rho,\theta)= \sum_{n,m} a_n^m
Z_n^m(\rho,\theta).
\end{equation}

In a turbulent atmosphere of Kolmogorov type, the aberrations are characterized by
coefficients $a_j$ that follow a multivariate Gaussian statistical distributions whose covariance
matrix  can be obtained analytically \citep{noll76}. It is interesting
to point out that the covariance matrix of the coefficients is not diagonal, although
it is possible to find modifications of the Zernike functions that diagonalize it.
%the
% covariance matrix \citep{roddier90}. Such modified functions are obtained using the Karhunen-Lo\`eve 
% transformation \citep[KL;][]{loeve55}, which finds the principal components through the
% diagonalization of the covariance matrix. In order to exploit the analytical properties
% of the Zernike functions, we keep using them as a good representation of the phase aberration.
% However, since the KL functions are just linear combinations of the Zernike functions,
% it would be possible to generalize the results presented below to the KL functions if the unitary transformation
% matrix is kept during the calculations.
Returning to the expression of the exponential of the phase aberration, we expand
it in power series:
\begin{equation}
e^{i \Phi(\rho,\theta)}=1+i\Phi(\rho,\theta)-\frac{1}{2}\Phi^2(\rho,\theta)+\ldots.
\end{equation} 
The radius of convergence of such expansion is infinite and that does not
translate into an easy cutoff of the series. But using the
decomposition in terms of Zernike functions of the phase aberration, we end up with:
\begin{equation}
e^{i \Phi(\rho,\theta)}=1+i\sum a_j Z_j(\rho,\theta)-\frac{1}{2}\left[\sum a_j
Z_j(\rho,\theta)\right]^2+\ldots
\label{eq:series_expansion}
\end{equation} 
For small aberrations ($a_j \ll 1$) the previous series can be safely cut at first order, something
that has been successfully exploited recently by \cite{janssen_enz02} and \cite{braat_enz02} to empirically estimate
point spread functions of microscope-type optical systems. Such approximation may be useful if one is interested in
describing aberrations introduced by well adjusted optical systems, but it is hardly
acceptable for aberrations introduced by atmospheric turbulence. Because of the complexity of the
resulting expression, there has been no interest in the past for this approach as a means to estimate the
PSF of atmospheric seeing. The reason is that the series expansion of 
Eq. (\ref{eq:series_expansion}) generates products $Z_j Z_k$ at second 
order and high-order products $Z_j \cdots Z_k$ at higher orders. In the absence of any clear rule to
manipulate such products, the power expansion was rendered useless already at
second order.

A recent mathematical result by \cite{Mathar09}\footnote{See also 
http://arxiv.org/abs/0809.2368} has radically simplified the 
approach to that expansion and we use those results here for the first time to pursue the
analytical integration of the PSF of \textit{any} aberration introduced by
atmospheric turbulence. \cite{Mathar09} constructively demonstrated that the product of two Zernike functions
can be written as a linear combination of Zernike functions (see Appendix \ref{sec:appendix_mathar}), so that:
\begin{equation}
\label{eq:expansion_mathar}
Z_j(\rho,\theta)Z_{j'}(\rho,\theta)=\sum_k d_k  Z_k(\rho,\theta),
\end{equation}
where the coefficients $d_k$ can be obtained by direct application of the relations
shown in Appendix \ref{sec:appendix_mathar}.
The product expansion of \cite{Mathar09} can be applied recursively without
difficulties, resulting in the following general result:
\begin{equation}
\prod_{j} Z_j(\rho,\theta)=\sum_k c_k  Z_k(\rho,\theta),
\end{equation}
where the $c_k$ coefficients can be calculated from the 
$d_k$ coefficients of Eq. (\ref{eq:expansion_mathar}).
As a consequence, the expansion of Eq. (\ref{eq:series_expansion}) can be written certainly as:
\begin{equation}
\label{eq:phase_aberration_expansion}
e^{i \Phi(\rho,\theta)}=1+\sum_{k=2} \beta_k  Z_k(\rho,\theta),
\end{equation}
where the $\beta_k$ coefficients are complex in general.
In other words, if the phase aberration can be expanded in Zernike functions, the wavefront can also be 
expanded efficiently in these functions with complex coefficients $\beta_k$ that can be inferred from the
coefficients $a_j$ by a repetitive application of the coupling relation of Eq. (\ref{eq:coupling})
and the coupling coefficient of Eq. (\ref{eq:coupling_coefficient}). A direct consequence of 
Eq. (\ref{eq:phase_aberration_expansion}) and Eq. (\ref{eq:series_expansion}) is that, for weak turbulence, the
imaginary parts of the $\beta$ coefficients are approximately equal to the Zernike coefficients
of the expansion of the wavefront:
\begin{equation}
\mathrm{Im}(\beta_k) \approx \alpha_k,
\end{equation}
and strictly equal at first order. The weak turbulence condition for that approximation often applies to our
astronomical applications as demonstrated statistically in Section \ref{sec:statistical}.
Through this approximation any method that allows us to estimate the value of the $\beta$ coefficients from the PSF
(like the phase diversity method shown below) directly gives us an estimation of the Zernike 
coefficients of the wavefront, at least to first order.

% \textsc{Since we have assumed that the pupil amplitude is equal to one and}
% the phase aberration function is real, the $\beta_k$ coefficients 
% are such that the modulus of the summation amounts to zero. Using the orthogonality of 
% the Zernike functions, the $\beta_k$ coefficients of the previous expansion can also be obtained by
% calculating the following integral:
% \begin{equation}
% \beta_j = 
% \frac{\gamma (n+1)}{\pi} \int_0^1\int_0^{2\pi} \rho d\rho d\theta Z_j(\rho,\theta) e^{i \Phi(\rho,\theta)} - \gamma \delta_{j1},
% \end{equation}
% where $\gamma=1$ for $m=0$ and $\gamma=2$ otherwise.

\begin{figure*}[!t]
\centering
\includegraphics[width=0.8\textwidth]{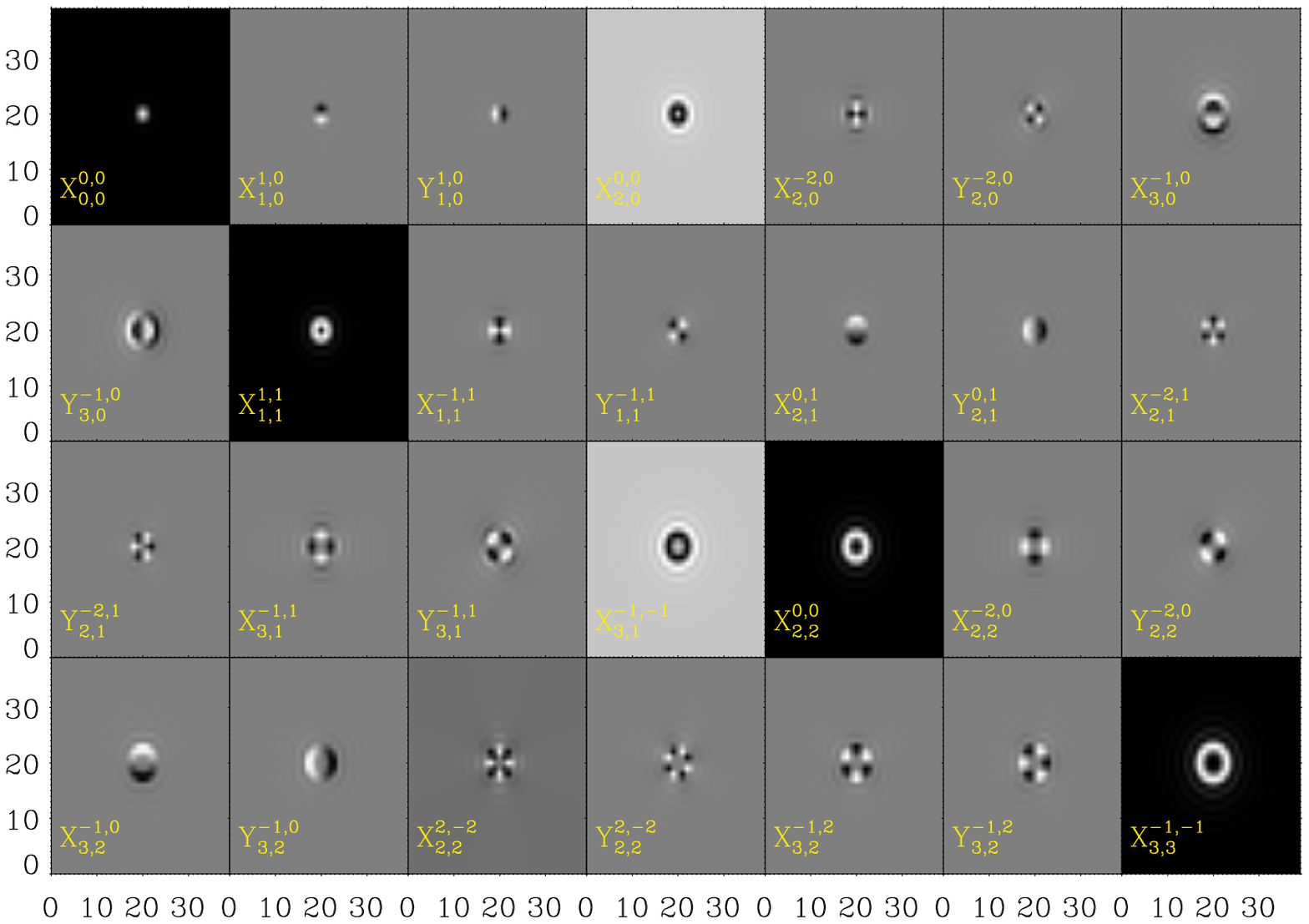}\\
\includegraphics[width=0.8\textwidth]{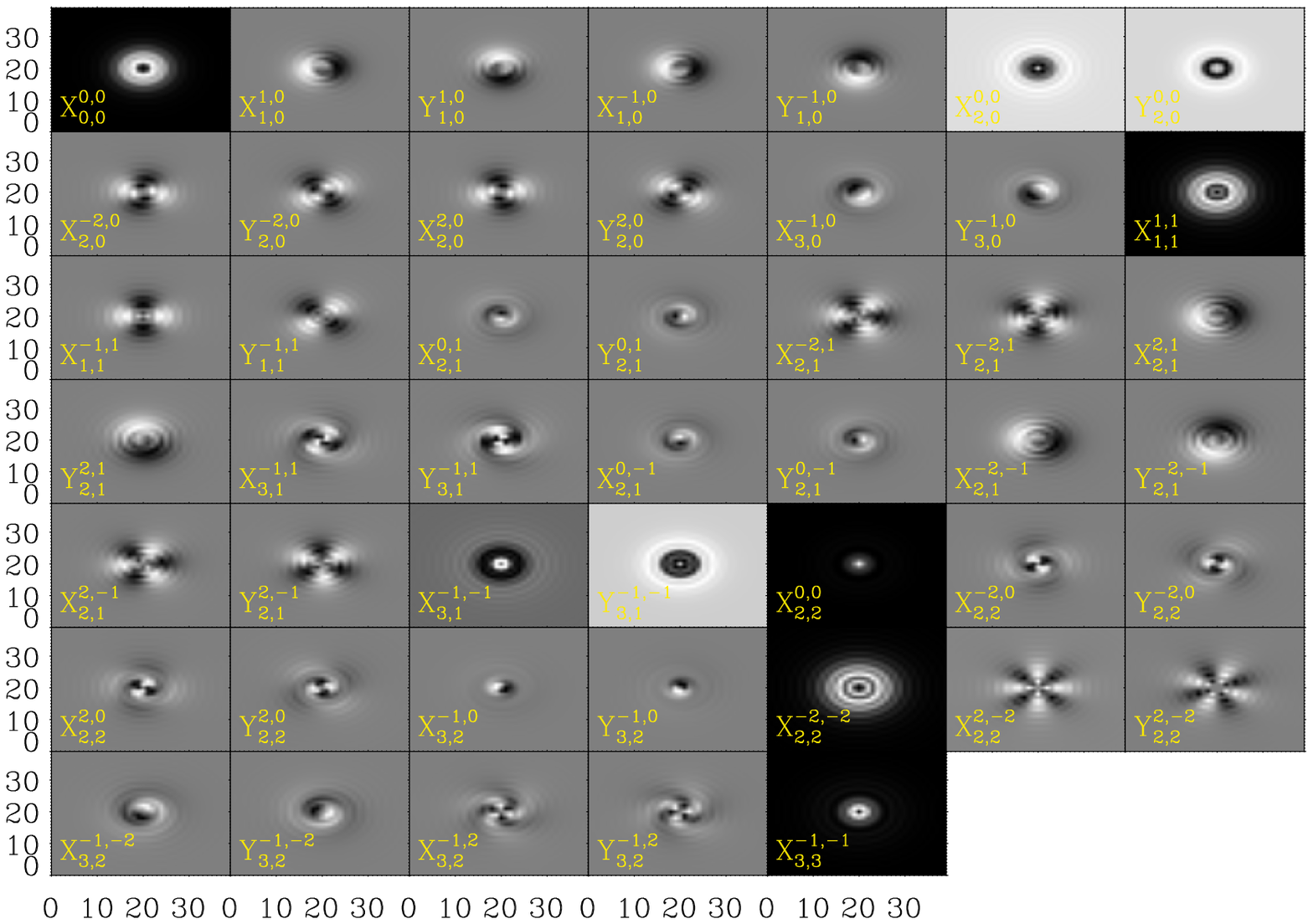}
\caption{Example of unique basis functions for representing the PSF including terms up to
the Noll index $j=7$. These functions are calculated for a telescope of 100 cm diameter at a wavelength
of 5250 \AA, a typical situation in present day telescopes. The spatial dimensions are in units of 
pixels, that we choose of size 0.055'', also typical in present observing conditions. The upper panel 
shows the results for the focused image while the lower panel shows the 
results for the defocused image, where the defocus is chosen as shown in \S\ref{sec:phase_diversity}.}
\label{fig:basis_set_psf}
\end{figure*}

%%%%%%%%%%%%%%%%%%%%%%%%%%
%%%%%%%%%%%%%%%%%%%%%%%%%%
\subsection{The field distribution}
Having written the pupil function $P(\rho,\theta)$ as a linear combination 
of Zernike functions with complex coefficients $\beta_k$, the
field distribution of Eq. (\ref{eq:fresnel}) simplifies to a sum of integrals of
the form:
% \begin{equation}
% U(r,\varphi;f)=\frac{1}{\pi}\int_0^1\int_0^{2\pi} \rho d\rho d\theta
% e^{if\rho^2} \left(1+\sum_j \beta_j Z_j\right) e^{2\pi i\rho r
% \cos(\theta-\varphi)}
% \end{equation}
\begin{eqnarray}
\label{eq:U_general}
U(r,\varphi;f)&=&\frac{1}{\pi}\int_0^1\int_0^{2\pi} \rho d\rho d\theta
e^{if\rho^2} e^{2\pi i\rho r \cos(\theta-\varphi)} \nonumber \\
&+&\sum_j \beta_j \frac{1}{\pi}\int_0^1\int_0^{2\pi} \rho d\rho d\theta
e^{if\rho^2} Z_j e^{2\pi i\rho r \cos(\theta-\varphi)}
\end{eqnarray}  
In the usual mathematical development of the ENZ theory we uncouple the two integrals above in radial and azimuthal parts. The azimuthal
part can be simplified using the Jacobi-Anger identity \citep{abramowitz72}:
\begin{equation}
\int_0^{2\pi} e^{2\pi i\rho r\cos(\theta-\varphi)}e^{im\theta} d\theta = 2\pi i
e^{im\varphi}J_m(2\pi \rho r),
\end{equation}
where the $J_m(2\pi \rho r)$ are Bessel functions of the first kind. Making use twice
of the previous integral identity in Eq. (\ref{eq:U_general}) and reformulating the
azimuthal part of the Zernike functions into a single $e^{im\theta}$ (that
contains both the $\cos$ and $\sin$ functions into
its real and complex parts for an unsigned $m$ index), we end up with:
\begin{eqnarray}
U(r,\varphi;f)&=&
2 \int_0^1 \rho d\rho  e^{if\rho^2} J_0(2\pi\rho r) \nonumber \\
&+&2\sum_j i^m e^{im\varphi} \beta_j V_n^m(r,f).
\end{eqnarray}
The function $V_n^m(r,f)$ represents the following (complex) radial integral:
\begin{equation}
\label{eq:V_n_m_definition}
V_n^m(r,f) = \int_0^1 \rho d\rho  e^{if\rho^2}  R_n^{|m|}(\rho)
J_m(2\pi\rho r),
\end{equation}
which fulfills that:
\begin{equation}
[V_n^m(r,f)]^\dag = V_n^m(r,-f).
\end{equation}
If we are interested in the field distribution at the plane where the image
would form in the absence of any optical aberration ($f=0$), the radial integral 
becomes real and has an easy solution:
\begin{equation}
V_n^m(r) = (-1)^{\frac{n-|m|}{2}}\frac{J_{n+1}(2\pi r)}{2\pi r}.
\end{equation}
Prefiguring its application in phase diversity, the integral of Eq. (\ref{eq:V_n_m_definition}) can be
numerically calculated for arbitrary values of $f$ using the rapidly convergent series expansion developed by
\cite{braat_enz02}.

The final expression for the field distribution is given by:
\begin{equation}
\label{eq:phase}
U(r,\varphi;f) = 2 V_0^0(r,f)+2\sum_j i^m e^{im\varphi}\beta_j V_n^m(r,f).
\end{equation}
Notice that, when no aberrations are present, the in-focus field distribution (at $f=0$) is proportional to:
\begin{equation}
V_0^0(r) = \frac{J_{1}(2\pi r)}{2\pi r}
\end{equation}
which is the function giving rise to the well known Airy spot for a circular
non-aberrated aperture. At $f \neq 0$, the field distribution is proportional to $V_0^0(r,f)$ which,
using the rapidly convergent series of \cite{braat_enz02}, can be written exactly as:
\begin{equation}
V_0^0(r,f) = e^{if} \sum_{l=1}^\infty (-2if)^{l-1} \frac{J_{l}(2\pi r)}{(2\pi r)^l}
\end{equation}

Summarizing,  the product expansion of
\cite{Mathar09} has allowed us to write the field
distribution in the image space as a linear combination of
functions $V_n^m(r,f)$  defined above, and of which the (in-focus or out-of-focus) 
Airy spot is just the first
contribution. This can be considered to be a generalization of the results presented
by \cite{janssen_enz02} and \cite{braat_enz02} for the case of small aberrations.

%%%%%%%%%%%%%%%%%%%%%%%%%%
%%%%%%%%%%%%%%%%%%%%%%%%%%
\section{The point spread function}
\label{sec:psf}
The point spread function or intensity distribution is obtained from the complex field distribution as:
\begin{equation}
\mathrm{PSF}(r,\varphi;f)= U(r,\varphi;f)U^\dag (r,\varphi;f).
\end{equation}
For the sake of simplicity in the presentation, we separate the analysis of the 
in-focus and the out-of-focus PSFs.

\subsection{In-focus PSF}
Using Eq. (\ref{eq:phase}) with the first term included in the summation as $j=0$, the PSF results in the
following double summation:
\begin{equation}
\mathrm{PSF}(r,\varphi)=4\sum_{j,k} i^m e^{im\varphi}\beta_j (-i)^{m'} e^{-im'\varphi}\beta_k^\dag V_n^m (r) V_{n'}^{m'}(r),
\end{equation}
where the $(n,m)$ pair is associated with the Noll index $j$ while the $(n',m')$ pair
is attached to the Noll index $k$.
The previous double summation can be separated into two summations: one containing
terms for which $j=k$ and one containing terms $j \neq k$. After tedious but straightforward
algebra, the final expression for the point spread function is:
\begin{eqnarray}
\label{eq:PSF_final}
% PSF(r,\varphi)&=&4 \sum_{j=0} |\beta_j|^2 \left[ V_n^m(r) \right]^2 \nonumber \\
% &+&8\sum_{\substack{j=0\\k<j}} \left[ C_{jk} X_{n,n'}^{m,m'}(r,\varphi) + D_{jk} Y_{n,n'}^{m,m'}(r,\varphi) \right],
\mathrm{PSF}(r,\varphi)=8\sum_{\substack{j=0\\k \leq j}} \left[ C_{jk} X_{n,n'}^{m,m'}(r,\varphi) + D_{jk} Y_{n,n'}^{m,m'}(r,\varphi) \right].
\end{eqnarray}
The functions appearing in the linear expansion can be calculated as:
\begin{eqnarray}
\label{eq:X_Y_zerofocus}
X_{n,n'}^{m,m'}(r,\varphi) &=& (-1)^{m'} C^{m,m'}(\varphi) V_n^m(r) V_{n'}^{m'}(r) \nonumber \\
Y_{n,n'}^{m,m'}(r,\varphi) &=& -(-1)^{m'} S^{m,m'}(\varphi) V_n^m(r) V_{n'}^{m'}(r).
\end{eqnarray}
Note that the azimuthal and radial contribution are separated. The angular dependence
is modulated by the following functions:
\begin{eqnarray}
\label{eq:trigonometric}
C^{m,m'}(\varphi) = \cos \left[ \frac{\pi}{2} (m+m')+(m-m')\varphi \right]  \nonumber \\
S^{m,m'}(\varphi) = \sin \left[ \frac{\pi}{2} (m+m')+(m-m')\varphi \right].
\end{eqnarray}

The coefficients $C_{jk}$ and $D_{jk}$ in the linear expansion of the PSF 
are related to the real and imaginary parts of products of $\beta$ coefficients:
\begin{eqnarray}
C_{jk} &=& C_{kj} = \epsilon_{jk} \mathrm{Re}(\beta_j \beta_k^\dag) \nonumber \\
D_{jk} &=& -D_{kj} = \epsilon_{jk} \mathrm{Im}(\beta_j \beta_k^\dag),
\end{eqnarray}
where $\epsilon_{kj}=1-\delta_{kj}/2$. As an example, Fig. \ref{fig:basis_set_psf} shows the 
unique basis functions up to Noll index $j=7$. The rest of functions with
different combinations of indices can be related to these by trivial relations like those shown in 
Appendix \ref{sec:properties_functions}. Note that the terms $j=k$ of the summation contain
only the contribution of the $C_{jk}$ coefficients and are azimuth-independent, since:
\begin{align}
X_{n,n}^{m,m}(r,\varphi) &= \left[ V_n^m(r) \right]^2 \nonumber \\
Y_{n,n}^{m,m}(r,\varphi) &= 0.
\end{align}

Because $n-|m|$ is even in the definition of the Zernike functions, it can be verified that only the
terms with $j=k$ in Eq. (\ref{eq:PSF_final}) contribute to the total area of the PSF, so that:
\begin{equation}
\label{eq:PSF_normalization}
\int_0^1 \int_0^{2\pi} r \mathrm{d}r \mathrm{d}\varphi \mathrm{PSF}(r,\varphi) = \sum_{j=0} \frac{|\beta_j|^2}{\pi(n+1)},
\end{equation}
where the value of $n$ is associated to the Noll index of summation.
As a result, the Strehl ratio defined as the ratio between the peak intensity at the origin of the PSF
and the equivalent Airy function simplifies to:
\begin{equation}
S = |\beta_0|^2 \left( \sum_{j=0} \frac{|\beta_j|^2}{(n+1)} \right)^{-1}.
\end{equation}

% The functions appearing in Eq. (\ref{eq:PSF_final}) are not orthogonal

\begin{figure*}[!t]
\centering
\includegraphics[width=0.8\textwidth]{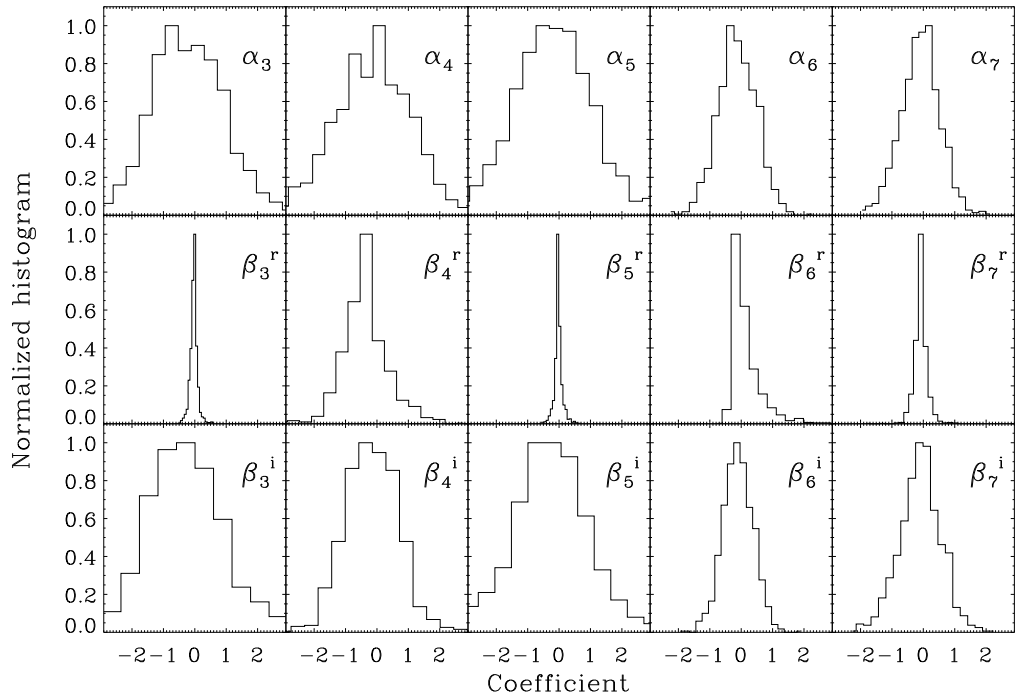}
\caption{Histograms of the distribution of the wavefront expansion coefficients $\alpha$ for an atmosphere
with Kolmogorov-type turbulence (upper row). The real and imaginary part of the corresponding expansion 
coefficients $\beta$ of the field distribution are shown in the middle and lower rows, respectively. The simulations are
carried out for a telescope of $D=90$ cm and a Fried parameter $r_0=7$ cm.}
\label{fig:histogram_betas}
\end{figure*}

%%%%%%%%%%%%%%%%%%%%%%%%%%
%%%%%%%%%%%%%%%%%%%%%%%%%%
\subsection{Out-of-focus PSF}
Following the same approach as in the previous section, we can also write an expression for
the PSF at any arbitrary focal position around the focal point. Since in this case the 
$V_n^m(r,f)$ functions are complex in general, the basis functions are slightly more elaborate:
\begin{align}
\label{eq:PSF_final_defocused}
% PSF(r,\varphi;f)&=4 \sum_{j=0} |\beta_j|^2 V_n^m {V_n^m}^\dag \nonumber \\
% &+8\sum_{\substack{j=0\\k<j}} \left[ C_{jk} X_{n,n'}^{m,m'}(r,\varphi;f) + D_{jk} Y_{n,n'}^{m,m'}(r,\varphi;f) \right],
\mathrm{PSF}(r,\varphi;f)=8\sum_{\substack{j=0\\k \leq j}} \left[ C_{jk} X_{n,n'}^{m,m'}(r,\varphi;f) + D_{jk} Y_{n,n'}^{m,m'}(r,\varphi;f) \right],
\end{align}
where
\begin{align}
% X_{n,n'}^{m,m'}(r,\varphi;f) &= (-1)^{m'} \Bigg\{ \nonumber \\
% &\cos \left[ \frac{\pi}{2} (m+m')+(m-m')\varphi \right] \xi_{n,n'}^{m,m'}(r;f) \nonumber \\
% &+\sin \left[ \frac{\pi}{2} (m+m')+(m-m')\varphi \right] \psi_{n,n'}^{m,m'}(r;f) \Bigg\} \nonumber \\
% Y_{n,n'}^{m,m'}(r,\varphi;f) &= (-1)^{m'} \Bigg\{ \nonumber \\
% &\cos \left[ \frac{\pi}{2} (m+m')+(m-m')\varphi \right] \psi_{n,n'}^{m,m'}(r;f) \nonumber \\
% &-\sin \left[ \frac{\pi}{2} (m+m')+(m-m')\varphi \right] \xi_{n,n'}^{m,m'}(r;f) \Bigg\}.
X_{n,n'}^{m,m'}&(r,\varphi;f) = (-1)^{m'} \nonumber \\
&\times \left[ C^{m,m'}(\varphi) \xi_{n,n'}^{m,m'}(r;f) +S^{m,m'}(\varphi) \psi_{n,n'}^{m,m'}(r;f) \right] \nonumber \\
Y_{n,n'}^{m,m'}&(r,\varphi;f) = (-1)^{m'} \nonumber \\
&\times \left[ C^{m,m'}(\varphi) \psi_{n,n'}^{m,m'}(r;f) -S^{m,m'}(\varphi) \xi_{n,n'}^{m,m'}(r;f) \right].
\label{eq:X_Y}
\end{align}
The auxiliary functions appearing in the previous formulae are given by:
\begin{align}
\xi_{n,n'}^{m,m'}(r;f) &= \mathrm{Im}(V_n^m) \mathrm{Im}(V_{n'}^{m'}) + \mathrm{Re}(V_n^m) \mathrm{Re}(V_{n'}^{m'}) \nonumber \\
\psi_{n,n'}^{m,m'}(r;f) &= \mathrm{Re}(V_n^m) \mathrm{Im}(V_{n'}^{m'}) - \mathrm{Im}(V_n^m) \mathrm{Re}(V_{n'}^{m'}).
\end{align}

Obviously, when the previous expressions are evaluated at the focus ($f=0$), because the $V_n^m$ 
functions are real, $\psi_{n,n'}^{m,m'}=0$ and $\xi_{n,n'}^{m,m'}=V_n^m V_{n'}^{m'}$, thus recovering 
the expressions of Eq. (\ref{eq:X_Y_zerofocus}). Likewise, for the terms of the summation in 
Eq. (\ref{eq:PSF_final_defocused}) for which $j=k$, we have:
\begin{align}
X_{n,n}^{m,m}(r,\varphi;f) &= \xi_{n,n}^{m,m}(r;f) = V_n^m(r;f) [{V_n^m(r;f)}]^\dag \nonumber \\
Y_{n,n}^{m,m}(r,\varphi;f) &= 0.
\end{align}

%%%%%%%%%%%%%%%%%%%%%%%%%%
%%%%%%%%%%%%%%%%%%%%%%%%%%
\section{Statistical properties of the \texorpdfstring{$\beta$}{beta} coefficients}
\label{sec:statistical}
In a Kolmogorov turbulent atmosphere, the coefficients $\alpha_j$ of the Zernike expansion of the
wavefront follow a multivariate Gaussian distribution with a non-diagonal covariance matrix determined 
by \cite{noll76}, \cite{wang_markey78} and \cite{roddier90}. The amplitude of the elements of the 
covariance matrix is determined by the Fried parameter $r_0$. As a function of this parameter, we 
simulate random samples $\alpha_k$ from such
multivariate Gaussian and use the expressions of Section \ref{sec:wavefront_expansion} and the
formulae developed by \cite{Mathar09} presented in Appendix \ref{sec:appendix_mathar} to calculate the 
equivalent $\beta_k$ coefficients. The corresponding statistical analysis is presented in
Fig. \ref{fig:histogram_betas} for aberrations of order higher than two, excluding tip-tilt. The upper
row shows that the $\alpha_k$ coefficients are Gaussian distributed, with a variance depending
on the order of the aberration. The middle and lower rows show the histograms for the real and
imaginary part of the $\beta_k$ coefficients, respectively. 

The distribution of the imaginary part is close to Gaussian and very similar to that of the 
wavefront expansion coefficients. As already mentioned, this is a consequence of the fact that,
to first order, the imaginary part of the $\beta_k$ coefficients equals the $\alpha_k$
coefficients. Higher-order corrections to the imaginary parts occur only at odd orders, so that
the first correction takes place already at third-order, which often induces a relatively 
small change for turbulence not too strong.
The Pearson linear correlation coefficient between $\alpha_k$ and $\mathrm{Im}(\beta_k)$ is larger
than 0.95 for practically all orders, indicating that $\mathrm{Im}(\beta_k)$ represents a good estimation
of the wavefront expansion coefficients. The real parts present distributions with a smaller variance and, in some cases, a 
large kurtosis. This is a consequence of the fact that the real part of the $\beta_k$ coefficients
is modified only at second-order often representing a small correction.

%%%%%%%%%%%%%%%%%%%%%%%%%%
%%%%%%%%%%%%%%%%%%%%%%%%%%
\section{Phase diversity}
\label{sec:phase_diversity}
A fundamental advantage of the analytical expression for the PSF 
just presented is that it allows to introduce a defocus in the PSF
while maintaining unchanged the $\beta_j$ coefficients that characterize the aberrations. As a consequence,
it is natural to propose such a functional form for doing image reconstruction first through phase
diversity techniques \citep{gonsalves79,paxman92,lofdahl_scharmer94}. 
% since the diversity introduced
% typically is a defocus. 
In order to fix naming conventions, we first summarize the basics of the standard approach to phase diversity.
\subsection{Standard approach}
The idea behind phase diversity is to simultaneously 
acquire a focused image ($D_0$) and another one with exactly the same aberrations plus a known 
defocus ($D_k$). Under the standard image formation paradigm, the observed image after degradation
by the atmosphere and the optical devices of the telescope can be written as:
\begin{eqnarray}
D_0 &=& F * P_0 + N_0 \nonumber \\
D_k &=& F * P_k + N_k,
\end{eqnarray}
where $P_0$ and $P_k$ are the PSFs of the atmosphere+telescope in the focused and defocused images, respectively.
The quantities $N_0$ and $N_k$ are noise contributions in the image formation, each one
characterized with a variance $\sigma_0^2$ and $\sigma_k^2$, respectively. Since both images are obtained
simultaneously, the underlying object $F$ is the same in both cases.

In the standard phase diversity method \citep[see, e.g.,][]{gonsalves79,paxman92,lofdahl_scharmer94}, the phase aberration 
in the focused image is expanded in a Zernike basis according to Eq. (\ref{eq:phase_aberration})\footnote{Some
advantage is gained by using instead the Karhunen-Lo\`eve expansion. The reason is that these functions are 
empirically built to explain the largest amount of turbulence variance with the fewer number of
functions.}. The PSFs are obtained from this phase aberration by calculating:
\begin{eqnarray}
P_0 &=& |\mathrm{FT}^{-1}\left\{A(\rho,\theta) \exp\left[{i \Phi(\rho,\theta)}\right]\right\}|^2 \nonumber \\
P_k &=& |\mathrm{FT}^{-1}\left\{A(\rho,\theta) \exp\left[{i \Phi(\rho,\theta)+i\delta_d(\rho,\theta)}\right]\right\}|^2,
\end{eqnarray}
where the defocus is introduced by adding a term of the form $\delta_d = \alpha^d_4 Z_4(\rho,\theta)$
and $\mathrm{FT}^{-1}$ stands for the inverse Fourier transform.
The election of the defocusing distance is essential for obtaining an efficient and successful phase
diversity method. It is usually chosen to represent one wave retardance at the wavelength of the observation.

% The rms defocus coefficient $a_4^d$ in radians is related to the
% defocusing distance $d$ by the relation:
% \begin{equation}
% a_4^d = \frac{\pi d}{8 \sqrt{3} \lambda (F/D)^2},
% \end{equation}
% with a peak-to-valley value of the wavefront error equal to:
% \begin{equation}
% \Delta = \frac{\sqrt{3} \lambda a_4^d}{\pi} = \frac{d}{8 (F/D)^2}.
% \end{equation}
% For instance, for 6000 \AA\ and a telescope like TH\'EMIS, the defocusing should be close to
% $d \approx 1.34$ mm, while for the examples we present in this paper carried out with
% the Swedish 1-m Solar Telescope \citep{scharmer02}, we get $d \approx 9$ mm for
% $\lambda=3954$ \AA. In our case, this optimal defocusing of one wavelength translates into a value of 
% the defocusing parameter of $f=2 \pi$.

The coefficients of the expansion in Zernike functions of the phase aberration are obtained by
calculating the coefficients $\alpha_j$ of the expansion of Eq. (\ref{eq:phase_aberration}) that
minimize the following error metric \citep{paxman92}:
\begin{equation}
L = \sum_{u,v} \frac{|\hat{D}_k \hat{P}_0 - \hat{D}_0 \hat{P}_k|^2}{(|\hat{P}_0|^2+\gamma |\hat{P}_k|^2)^{1/2}},
\label{eq:metric}
\end{equation}
where $\hat{D}_0$, $\hat{D}_k$, $\hat{P}_0$ and $\hat{P}_k$ are the Fourier transforms of $D_0$, $D_k$, $P_0$ and $P_k$, 
respectively. The summation is carried out over the full Fourier domain defined by
the frequencies $u$ and $v$, while $\gamma=\sigma_0^2/\sigma_k^2$.
Note that the dependence of the metric on the $\alpha_j$ coefficients is highly non-linear due to the
exponential function, the Fourier transforms and the modulus operation that appears in the
definition of the PSF. Note also that each
evaluation of the metric and its gradient with respect to the $a_j$ coefficients needs to 
carry out several Fourier transforms, which are of order $O(N \log N)$. At a final step, once
the PSF (and the ensuing Fourier transform) are known, the reconstructed image is estimated
as the inverse Fourier transform of \citep{paxman92}:
\begin{equation}
\label{eq:optimal_image}
\hat{F} = \frac{\hat{D}_0 \hat{P}_0^\dag + \gamma \hat{D}_k \hat{P}_k^\dag}{|\hat{P}_0|^2+\gamma |\hat{P}_k|^2}.
\end{equation}

The maximum-likelihood solution of the metric of \cite{gonsalves79} is not affected by the presence 
of additive Gaussian noise but it strongly affects the deconvolution
process given by Eq. (\ref{eq:optimal_image}) since noise is unlimitedly amplified. We follow 
\cite{lofdahl_scharmer94} and apply a filter $\hat H$ in the Fourier domain to the Fourier transform
of the observed images with the aim of filtering a large part of the noise. 
% Such filter is given
% by the expression:
% \begin{equation}
% \hat H = 1 - \langle |N|^2 \rangle \left \langle \frac{|\hat{P}_0|^2+\gamma |\hat{P}_k|^2}{\hat{D}_0 \hat{P}_0^\dag + \gamma \hat{D}_k \hat{P}_k^\dag},
% \right \rangle
% \end{equation}
% where $\langle |N|^2 \rangle$ is the noise expectation value. 
The same procedures applied by \cite{lofdahl_scharmer94}
for filtering remaining noise peaks at high frequency are also applied.

% Since we have only modified the description of the PSF with respect to previous
% approaches, all common reconstruction schemes (with any desired complexity) can be extended 
% to use our formalism. Among them, we find methods like multi-image phase diversity 
% \citep[e.g.,][]{paxman92} or multi-object multi-frame blind deconvolution \citep{vannoort05}
% that use many images to estimate information about frequencies that has been destroyed
% by the presence of noise. In essence, all these schemes require writing an error
% metric (equivalently, a likelihood function) like Eq. (\ref{eq:metric}) that takes into
% account the presence of the additional information. The larger amount of information
% helps regularize the problem and reduces the influence of noise. However, we also point out that
% it is possible to introduce regularization by using a fully Bayesian approach in which
% prior information is introduced in the problem.

\subsection{The new approach}
The crucial advantage 
of the phase diversity algorithm we propose is that the PSF of the focused and the defocused
images can be written as linear combination of known functions. Because of that, the previous error metric can
be minimized without performing a single Fourier transform during the iterative scheme. After
Fourier-transforming the observed images and the basis functions $X$ and $Y$, it is a matter of
finding the values of the $\beta_j$ complex quantities that minimize the metric $L$. Another advantage is
that the non-linear dependence of the metric function on the $\beta_j$ coefficients is less
pronounced, thus helping the optimization routine to find the minimum. 

In order to find the values of the coefficients $\beta_j$ that minimize the metric $L$, we
use a scaled conjugate gradient algorithm \citep{scalcg_andrei08} that performs extremely well
and fast for complex problems. To this end, one needs the gradient of $L$ with respect to the $\beta_j$
coefficients. After a tedious but straightforward algebra, the gradient can be found to be:
\begin{eqnarray}
\frac{\partial L}{\partial \beta_l^\mathrm{r,i}}&=&2\sum \mathrm{Re} \Bigg[ Q^2(\hat{D}_k \hat{P}_0 - \hat{D}_0 \hat{P}_k)^\dag 
\left( \hat{D}_k \frac{\partial \hat{P}_0}{\beta_l^\mathrm{r,i}} - 
\hat{D}_0 \frac{\partial \hat{P}_k}{\beta_l^\mathrm{r,i}} \right) \nonumber \\
&-&Q^4 |\hat{D}_k \hat{P}_0 - \hat{D}_0 \hat{P}_k|^2 \mathrm{Re} \left( \frac{\partial \hat{P}_0}{\beta_l^\mathrm{r,i}} \hat{P}_0^\dag + \gamma
\frac{\partial \hat{P}_k}{\beta_l^\mathrm{r,i}} \hat{P}_k^\dag \right) \Bigg]
\end{eqnarray}
where $Q=(|\hat{P}_0|^2+\gamma |\hat{P}_k|^2)^{-1/2}$ and:
\begin{eqnarray}
\frac{\partial \hat{P}_0}{\beta_l^\mathrm{r}} &=& 8 \sum_{k=0} \left( \beta_k^\mathrm{r} \hat{X}_{n,n'}^{m,m'}(r,\phi;0) - 
\beta_k^\mathrm{i} \hat{Y}_{n,n'}^{m,m'}(r,\phi;0)\right) \nonumber \\
\frac{\partial \hat{P}_0}{\beta_l^\mathrm{i}} &=& 8 \sum_{k=0} \left( \beta_k^\mathrm{r} \hat{Y}_{n,n'}^{m,m'}(r,\phi;0) +
\beta_k^\mathrm{i} \hat{X}_{n,n'}^{m,m'}(r,\phi;0)\right)\nonumber \\
\frac{\partial \hat{P}_k}{\beta_l^\mathrm{r}} &=& 8 \sum_{k=0} \left( \beta_k^\mathrm{r} \hat{X}_{n,n'}^{m,m'}(r,\phi;f) - 
\beta_k^\mathrm{i} \hat{Y}_{n,n'}^{m,m'}(r,\phi;f)\right) \nonumber \\
\frac{\partial \hat{P}_k}{\beta_l^\mathrm{i}} &=& 8 \sum_{k=0} \left( \beta_k^\mathrm{r} \hat{Y}_{n,n'}^{m,m'}(r,\phi;f) +
\beta_k^\mathrm{i} \hat{X}_{n,n'}^{m,m'}(r,\phi;f)\right).
\end{eqnarray}
The superindices r and i indicate the real and imaginary part of the coefficient $\beta_l$, respectively.
One should be aware that the values of $n$ and $m$ in the previous equation are those associated to the Noll
index $l$, while $n'$ and $m'$ are those associated to the index $k$ of the summation. Finally, the functions
$\hat{X}$ and $\hat{Y}$ are the Fourier transforms of the functions defined in Eq. (\ref{eq:X_Y}).

\begin{figure*}[!t]
\centering
\includegraphics[width=0.55\textwidth]{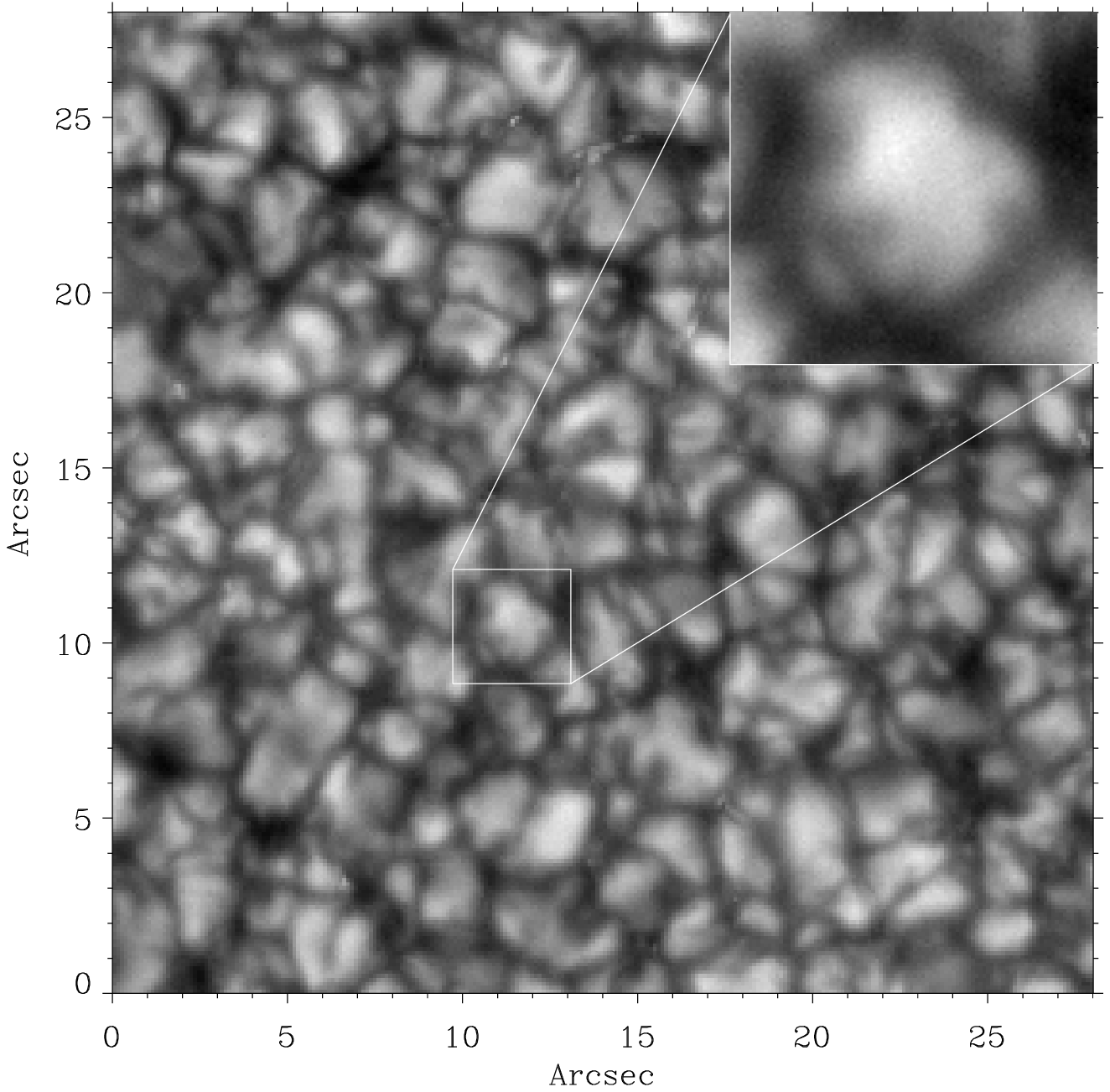}%
\includegraphics[width=0.55\textwidth]{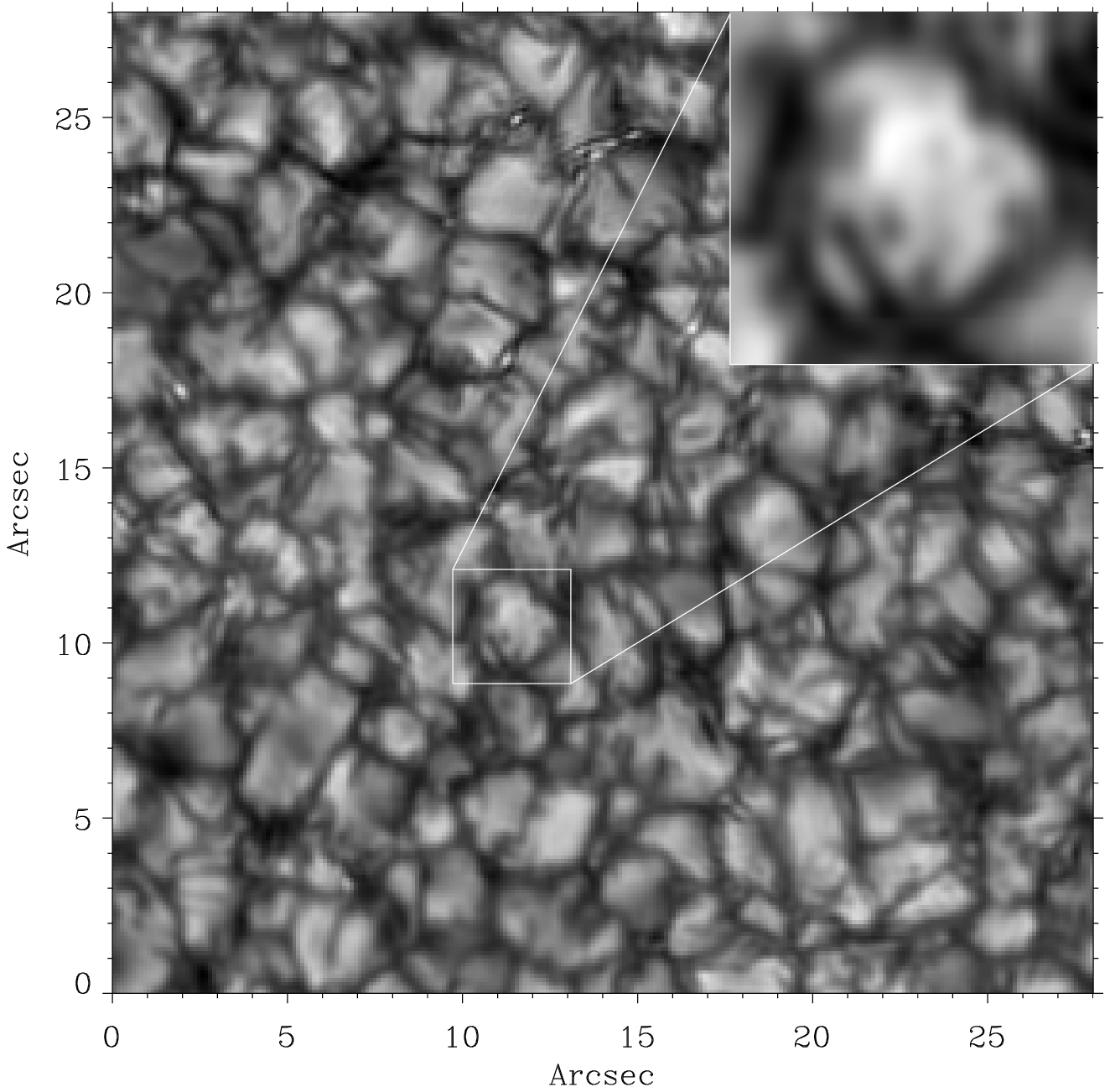}
\caption{Reconstruction example at 395 nm. The lower
image shows the in-focus image while the right image presents the reconstructed image using only one
pair of in-focus and defocused images. The contrast of the original image is 9.6\% while the reconstructed
one presents a contrast of 11.9\%. Data has been acquired with the SST telescope.}
\label{fig:pd_example_sst}
\end{figure*}

The computer code is written in Fortran 90 and runs in parallel environments using the Message Passing Interface library. Prior
to the phase diversity restoration procedure, there is an initial phase in which the basis functions of 
Eqs. (\ref{eq:X_Y_zerofocus}) and (\ref{eq:X_Y}) are precalculated and their Fourier transforms are saved 
for later use. Since this has to be run only once for each telescope and size of the isoplanatic patch, it
can be done with the required precision at virtually no extra computing time. In a second phase, in which the 
actual phase diversity restoration 
process is carried out, these Fourier transforms of the basis functions are quickly read and used in all 
subsequent calculations. This greatly accelerates the computing time and only two Fourier transforms are
needed per isoplanatic patch. The parallel implementation scales very well with the number of processors. 
The field-of-view is divided into isoplanatic patches of a given size and the reconstruction of each patch 
is carried out by one of the available processors. In principle, if $N_\mathrm{proc}$ are at hand, the computing
time is reduced by a factor $1/N_\mathrm{proc}$ because the contribution of all input/output tasks is negligible
with respect to the actual iterative process to minimize the phase diversity merit function.

\subsection{Illustrative examples}
We present two representative examples of the success of the implementation of our analytical
representation of the PSF for carrying out phase-diversity image reconstruction. We stress that our purpose
is not to present amazing reconstructions but to demonstrate that reconstruction can be
made to the same level of quality of the standard procedures only much faster.

\subsubsection{Blue images}
In our first example the quiet Sun internetwork was observed with the Swedish 1-m Solar Telescope
\citep[SST;][]{scharmer02} during 2007 September 29 during a period
of very good seeing \citep[see][for a description of the observations]{bonet08}. The pixel
size is $0.034''$ and the observations are carried out in the continuum close
to the Ca H line at 3953 \AA, providing a theoretical diffraction limit of $0.1''$. All these parameters were chosen to enhance granulation contrast.
With an isoplanatic patch of 128$\times$128 pixels ($\sim$4.5$''\times$4.5$''$), an image 
of 1000$\times$1000 pixels contains approximately
60 isoplanatic patches. Our phase diversity restoration algorithm takes around 5 s per patch when including
20 terms in the summation of Eq. (\ref{eq:phase_aberration_expansion}) that describes the wavefront. Therefore, the
total computing time is $\sim$5 min. When run in a standard present-day desktop computer with 4 processors, every
image can be fully restored in a timescale of 1 minute. Fig. \ref{fig:pd_example_sst} presents the
results of the reconstruction (right panel) and the original focused image (left panel). The
region zoomed illustrates the spatial frequencies enhanced by the phase-diversity algorithm.

\subsubsection{Red images}
\begin{figure}[!t]
\centering
\includegraphics[width=1.4\columnwidth]{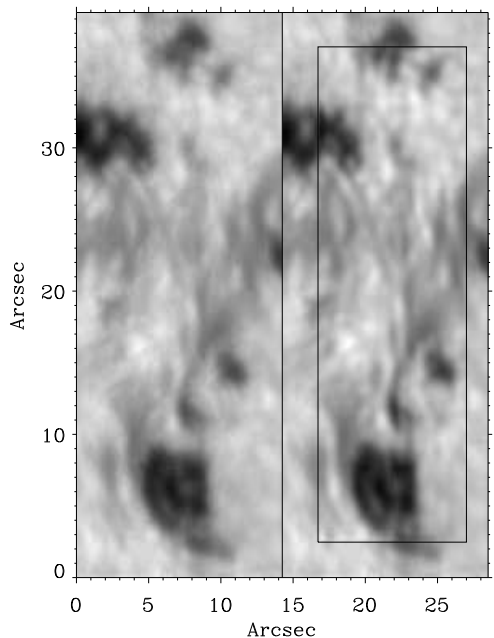}
\caption{Example of the phase-diversity reconstruction 
at 850 nm. We represent the focused image (left) and the reconstructed sub-image indicated 
with a box (right). Images have been acquired with THEMIS.}
\label{fig:pd_example_themis}
\end{figure}

In this second  example, we analyze observations of an active region carried out during 2002 May 15 with
the THEMIS telescope \citep{rayrole_mein93} at a wavelength of 850 nm. The pixel size is $0.075''$ and, due to
the large wavelength, the diffraction limit is $0.24''$, limiting the contrast of the granulation, both because 
of the wavelength and the intrinsically reduced contrast in the infrared. This dataset has been also
use by \cite{criscuoli05} to present the first results of phase-diversity applied
to THEMIS data. The images were restored with patches of 100$\times$100 pixels,
equivalent to 7.5$''\times$7.5$''$. Patches were finally mosaicked to
build the final image shown in Fig. \ref{fig:pd_example_themis}, were only
the part inside the rectangle has been reconstructed, using 20 terms in the 
summation of Eq. (\ref{eq:phase_aberration_expansion}) that describes the wavefront. Visual inspection
clearly indicates an improvement in the image quality. If desired, better visual and quantitative 
results can be obtained by adding multi-image information to the reconstruction algorithm
\citep{criscuoli05}.

Computing time is what makes the difference, since each patch, as in the previous case, only took 5 sec 
in a standard desktop computer. With such requirements, image reconstruction can be easily implemented as 
on line task in the data pipelines of the telescopes themselves.

%%%%%%%%%%%%%%%%%%%%%%%%%%
%%%%%%%%%%%%%%%%%%%%%%%%%%
\section{Conclusions}
\label{sec:conclusions}

The use of the recently developed Extended Nijboer-Zernike theory together with some mathematical results on 
the multiplication of Zernike polynomials has allowed us to rewrite the image formation 
integrals with an aberrated wavefront in terms of linear combinations of analytic functions. 
Generalizing the usual ENZ theory, we are able to do so   \textit{a priori} independently of 
the amplitude of the aberrations. With such mathematical tool we are able to rewrite the 
techniques of post-facto image reconstruction taking advantage primarily of the fact that 
the Fourier transforms of those analytic functions can be precomputed once and for all. 
Building different PSFs in the minimization process at the core of these reconstruction algorithms 
does not require to recompute any other Fourier transform but just modify the scalar, although 
complex, coefficients multiplying the functions.

The gain in speed is enormous. We illustrate it through two observations readied for phase 
diversity. Image reconstruction with quality identical to the standard algorithms is performed 
in question of seconds per patch, and one to five minutes for wide-field images, in standard desktop computers.
If image reconstruction is a must in present and future solar observations, the bottleneck of 
computationally expensive and time consuming algorithms was almost a showstopper for future 
instruments. The present improvement through the use of analytical PSFs solves the core of that problem and 
suggests that image reconstruction can even be implemented as a routine on-line procedure on the 
data pipelines of the telescope instruments themselves.

Since we have only modified the description of the PSF with respect to previous
approaches, all common reconstruction schemes (with any desired complexity), and not just phase diversity, can be extended 
to use our formalism. Among them, we find methods like multi-image phase diversity 
\citep[e.g.,][]{paxman92} or multi-object multi-frame blind deconvolution \citep{vannoort05}
that use many images to estimate information about frequencies that has been destroyed
by the presence of noise. In essence, all these schemes require writing an error
metric (equivalently, a likelihood function) like Eq. (\ref{eq:metric}) that takes into
account the presence of the additional information. The larger amount of information
helps regularize the problem and reduces the influence of noise. All of them are penalized by the 
computing time involved and all of them can
potentially take profit of the analytical approach we present to reduce the computational 
burden. The final goal is not to slow down the flows of
images from instruments because of the lack of capability to handle the image reconstruction problem.

Beyond those known techniques, we also want to point out that
it is possible to introduce regularization by using a fully Bayesian approach in which
prior information is introduced in the problem, eliminating a priori the noise nuisance
in the minimization of the metric.

\begin{acknowledgements}
Images from the Swedish Solar Tower (SST) were provided by J. A. Bonet (IAC). Images from THEMIS were provided by D. del Moro (Univ. Roma 
Tor Vergata). We are very thankful to both of them for providing the data necessary to illustrate this work. 
Financial support by the Spanish Ministry of Science and Innovation through project AYA2007-63881 is gratefully acknowledged.
\end{acknowledgements}

\begin{appendix}
\section{Product of Zernike functions}
\label{sec:appendix_mathar}
The breakthrough brought over by \cite{Mathar09} has been to point that, in the
product of two Zernike functions, only the following three situations could
arise:
\begin{eqnarray} 
2 \cos (m_1 \theta)\cos(m_2 \theta) &=&
\cos[(m_1-m_2)\theta] 
+ \cos[(m_1+m_2)\theta] \nonumber \\
2 \sin (m_1 \theta)\cos(m_2 \theta) &=&
\sin[(m_1-m_2)\theta]+\sin[(m_1+m_2)\theta] \nonumber \\
2 \sin (m_1 \theta)\sin(m_2 \theta) &=&
\cos[(m_1-m_2)\theta]-\cos[(m_1+m_2)\theta]. 
\end{eqnarray}
In other words, the result is always the sum of trigonometric functions depending on
two azimuthal numbers $m_1 \pm m_2$. This is indeed the seed of two new Zernike
functions with $m_3=m_1 \pm m_2$ if we succeed in writing the product of the
radial polynomials with identical $m_3$ azimuthal value. That is, we seek a
product expansion for the radial polynomials of the form:
\begin{equation}
R_{n_1}^{m_1} R_{n_2}^{m_2} = \sum _{n3=m3}^{n_1+n_2}
g_{n_1,m_1,n_2,m_2,n_3,m_3} R_{n_3}^{m_3}.
\label{eq:coupling}
\end{equation}
Every term of the series above has the same azimuthal number $m_3$,
coincident with the $m_3$ of the trigonometric product and can therefore be
combined with it to build a Zernike function $Z_{n_3}^{m_3}$. \cite{Mathar09}
demonstrates that this is indeed feasible and provides the value of the
$g_{n_1,m_1,n_2,m_2,n_3,m_3}$ coefficient in the product expansion as
\begin{eqnarray}
&&g_{n_1,m_1,n_2,m_2,n_3,m_3} = 2(n_3+1) \nonumber \\
&\times& \sum_{s_1=0}^{-a_1}\sum_{s_2=0}^{-a_2}\sum_{s_3=0}^{-a_3}
\frac{1}{n_1+n_2+n_3+2(1-s_1-s_2-s_3)} \nonumber \\
&\times& \prod_{j=1}^3 (-1)^{s_j}\binom{n_j-s_j}{s_j}\binom{n_j-2s_j}{-a_j-s_j},
\label{eq:coupling_coefficient}
\end{eqnarray}  
where $a=-(n-m)/2$. These coefficients play the same role of the Clebsh-Gordan recoupling coefficients
for the product of two spherical harmonics.

\section{Some properties of \texorpdfstring{$X_{n,n'}^{m,m'}$}{X(n,n')(m,m')} and \texorpdfstring{$Y_{n,n'}^{m,m'}$}{Y(n,n')(m,m')} and analytical Strehl ratio}
\label{sec:properties_functions}
Under the interchange of the $(n,m)$ and $(n',m')$ pair of indices, the following expressions hold:
\begin{eqnarray}
\xi_{n',n}^{m',m}(r;f) &=& \xi_{n,n'}^{m,m'}(r;f) \nonumber \\
\psi_{n',n}^{m',m}(r;f) &=& -\psi_{n,n'}^{m,m'}(r;f) \nonumber \\
C^{m',m}(\varphi) &=& C^{m,m'}(-\varphi) \nonumber \\
S^{m',m}(\varphi) &=& S^{m,m'}(-\varphi).
\end{eqnarray}
As a consequence, the functions $X_{n,n'}^{m,m'}$ and $Y_{n,n'}^{m,m'}$ fulfill the following
properties:
\begin{eqnarray}
X_{n',n}^{m',m}(r,\varphi;f) &=& (-1)^{m'-m} Y_{n,n'}^{m,m'}(r,-\varphi;f) \nonumber \\
Y_{n',n}^{m',m}(r,\varphi;f) &=& -(-1)^{m'-m} X_{n,n'}^{m,m'}(r,-\varphi;f).
\end{eqnarray}

Concerning the Strehl ratio it is possible to
demonstrate that only the Airy part of the PSF (either in-focus
or out-of-focus) contributes to the central peak of the PSF. The reason is that, evaluating Eq. (\ref{eq:V_n_m_definition}) at $r=0$
gives:
\begin{equation}
V_n^m(0,f) = \delta_{n0} \delta_{m0} \frac{1}{2f} \left[ \sin f + i(1- \cos f)\right],
\end{equation}
so that
\begin{eqnarray}
X_{n,n'}^{m,m'}(r,\varphi;f) &=& \delta_{n0} \delta_{m0} \delta_{n'0} \delta_{m'0} \frac{1-\cos f}{2f^2} \nonumber \\
Y_{n,n'}^{m,m'}(r,\varphi;f) &=& 0,
\end{eqnarray}
which gives, for the $f=0$, a contribution of 1/4 at the center of the PSF.
\end{appendix}

% \bibliographystyle{aa}
% \bibliography{/home/aasensio/Dropbox/biblio}

\end{document}